\documentclass[sigconf,screen, 9pt]{acmart}
\usepackage{subcaption}
\acmSubmissionID{97}
\usepackage{url}
\usepackage{enumitem}
\usepackage{graphicx}
\usepackage{caption}
\usepackage{amsmath}
\usepackage{tabu}
\usepackage{multirow}
\usepackage{hyperref}
\usepackage{adjustbox}
\usepackage{mathptmx}
\usepackage{amsfonts}
\usepackage{algpseudocode}
\usepackage{pgfplots}
\pgfplotsset{compat=1.12}
\usepackage{soul}
\usepackage{footnote}
\usepackage{lipsum}
\usepackage[ruled,vlined, linesnumbered]{algorithm2e}
\usepackage{subcaption}
\usepackage{listings}

\colorlet{punct}{red!60!black}
\definecolor{background}{HTML}{FFFFFF}
\definecolor{delim}{RGB}{20,105,176}
\colorlet{numb}{magenta!60!black}

\lstdefinelanguage{json}{
    basicstyle=\footnotesize\ttfamily,
    numbers=left,
    numberstyle=\scriptsize,
    xleftmargin=2.3em,
    xrightmargin=0.5em,
    framexleftmargin=1.9em,
    stepnumber=1,
    numbersep=8pt,
    showstringspaces=false,
    breaklines=true,
    frame=single,
    backgroundcolor=\color{background},
    literate=
     *{0}{{{\color{numb}0}}}{1}
      {1}{{{\color{numb}1}}}{1}
      {2}{{{\color{numb}2}}}{1}
      {3}{{{\color{numb}3}}}{1}
      {4}{{{\color{numb}4}}}{1}
      {5}{{{\color{numb}5}}}{1}
      {6}{{{\color{numb}6}}}{1}
      {7}{{{\color{numb}7}}}{1}
      {8}{{{\color{numb}8}}}{1}
      {9}{{{\color{numb}9}}}{1}
      {:}{{{\color{punct}{:}}}}{1}
      {,}{{{\color{punct}{,}}}}{1}
      {\{}{{{\color{delim}{\{}}}}{1}
      {\}}{{{\color{delim}{\}}}}}{1}
      {[}{{{\color{delim}{[}}}}{1}
      {]}{{{\color{delim}{]}}}}{1},
}

\AtBeginDocument{%
  \providecommand\BibTeX{{%
    \normalfont B\kern-0.5em{\scshape i\kern-0.25em b}\kern-0.8em\TeX}}}

\copyrightyear{2022} 
\acmYear{2022} 
\setcopyright{acmcopyright}\acmConference[WoSC '22]{Eighth International Workshop on Serverless Computing}{November 7, 2022}{Quebec, QC, Canada}
\acmBooktitle{Eighth International Workshop on Serverless Computing (WoSC '22), November 7, 2022, Quebec, QC, Canada}
\acmPrice{15.00}
\acmDOI{10.1145/3565382.3565881}
\acmISBN{978-1-4503-9927-2/22/11}

\newcommand{\shrinkspace}{\vspace{-5mm}}
\begin{document}

\title{Migrating from Microservices to Serverless: An IoT Platform Case Study}

\author{Mohak Chadha, Victor Pacyna, Anshul Jindal, Jianfeng Gu, Michael Gerndt}



\email{mohak.chadha@tum.de, victor.pacyna@tum.de, anshul.jindal@tum.de, jianfeng.gu@tum.de, gerndt@in.tum.de}
\affiliation{%
 \institution{Chair of Computer Architecture and Parallel Systems \\ Technische Universit{\"a}t M{\"u}nchen}
  \city{Garching (near Munich)}
  \state{Germany}
}

\renewcommand{\shortauthors}{Chadha et al.}

\begin{abstract}

Microservice architecture is the common choice for developing cloud applications these days since each individual microservice can be independently modified, replaced, and scaled. As a result, application development and operating cloud infrastructure were bundled together into what is now commonly called \textit{DevOps}. However, with the increasing popularity of the serverless computing paradigm and its several advantages such as no infrastructure management, a \texttt{pay-per-use} billing policy, and on-demand fine-grained autoscaling, there is a growing interest in utilizing  FaaS and serverless CaaS technologies for refactoring microservices-based applications. Towards this, we migrate a complex IoT platform application onto OpenWhisk (OW) and Google Cloud Run (GCR). We comprehensively evaluate the performance of the different deployment strategies, i.e.,  Google Kubernetes Engine (GKE)-Standard, OW, and GCR for the IoT platform using different load testing scenarios. Results from our experiments show that while GKE standard performs best for most scenarios, GCR is always cheaper wrt costs.

\end{abstract}



\maketitle

\section{Introduction}
\label{sec:intro}
Cloud computing is the corner-stone of modern day corporate and consumer IT. For companies it provides a multitude of advantages such as outsourcing of costly infrastructure, scalability, elasticity, and fault tolerance. To this end, the architectures of cloud native applications have adjusted to the ever increasing demand of migration to the cloud. Most prominent among them is the \textit{microservices} architecture that displaced the prevalent monolithic architecture~\cite{fowlermic}. This architecture enables the development of applications as a suite of small independent services communicating with each other via well-defined interfaces. A microservices-based architecture provides several advantages such as better modularization, enhanced isolation, and easier maintainability~\cite{newman2021}. Moreover, each service can be scaled up or down on-demand, or be deployed in multiple availability zones to eliminate single point of failures. However,  this has led to an increase in complexity of deployment and provisioning of services, resulting in developers having to develop their application as well as take care of its operation, i.e., DevOps~\cite{jindal2020}. To this end, an emerging computing paradigm called \textit{serveless computing} that abstracts infrastructure management away from the user has gained popularity and widespread adoption in various application domains such as edge computing~\cite{jindal2021function, fado, tinyfaas} and machine learning~\cite{fedless, Chadha2020}.


Function-as-a-Service (FaaS) and serverless Container-as-a-Service (CaaS) are key enablers of serverless computing that allow developers to focus on the application logic, while responsibilities such as infrastructure management, resource provisioning, and scaling are handled by the cloud service providers. In FaaS or serverless CaaS, the developer implements fine-grained  functions that are executed in response to external triggers such as HTTP requests and deploys them into a FaaS platform such as Apache OpenWhisk (OW)~\cite{openwhisk} or a serverless CaaS platform such as Google Cloud Run (GCR)~\cite{GCR}. On invocation, the platform creates an execution environment, i.e., \textit{function instance} which provides a secure and isolated language-specific \textit{runtime} environment for the function. These functions are generally launched on the platform's traditional IaaS virtual machine offerings~\cite{architecture, demystifying}. Serverless computing offers several advantages such as \texttt{scale-to-zero} for idle functions, a \texttt{pay-per-use} billing policy, and rapid fine-grained automatic scaling on a burst of function invocation requests. However, serverless functions are stateless and any application state needs to be maintained through external transactions with a database or a data store. 

Both architectures, i.e., either microservices or serverless have their advantages and disadvantages and the decision to adopt one over the other depends on several factors. In this paper, we investigate these two competing software architectures for an IoT platform application. Towards this, our key contributions are:
\begin{itemize}
    \item We migrate a microservices-based IoT platform application developed by us onto OW and GCR. 
    \item We comprehensively evaluate the performance of the IoT platform across different deployment strategies, i.e, Google Kubernetes Engine (GKE) standard, OW, and GCR with different load testing scenarios wrt performance and cost.
    \item We highlight the important lessons learned from our migration. All code artifacts related to this work are available at\footnote{https://github.com/CAPS-Cloud/IoT-Platform-Migration}.
\end{itemize}

The rest of the paper is structured as follows. \S\ref{sec:iotplatoverview} describes the IoT platform application in detail. In \S\ref{sec:methodology}, our migration methodology is described. \S\ref{sec:results} presents our experimental results. In \S\ref{sec:relatedwork}, prior work on migrating microservices onto FaaS platforms is described. Finally, \S\ref{sec:conclusion} concludes the paper and presents an outlook.

\section{IoT Platform}
\label{sec:iotplatoverview}


We designed and developed the IoT platform to enable numerous IoT applications such as room monitoring and occupancy estimation. The platform provides a scalable infrastructure for managing multiple users, devices, and sensors. Furthermore, it provides persistent storage for sensor data that is received via \texttt{HTTP}, \texttt{WebSocket} or \texttt{MQTT} protocols and supports secure communication with end devices through \texttt{JWT} tokens. Finally, it supports visualization of data in various formats to enable smart data analytics.


\subsection{Data Model}
\label{sec:datamodel}

\begin{figure}[t]
\centering
\includegraphics[width=0.9\columnwidth]{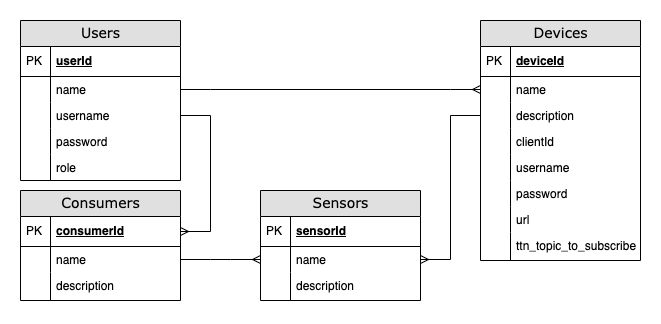}
\caption{Data Model of the IoT platform.} 
\label{fig:iotplatformdata}
\shrinkspace
\end{figure}

The IoT platform's data model consists of four main entities, i.e., \textit{users}, \textit{devices}, \textit{sensors}, and \textit{consumers}. Users are the main components responsible for interacting with the IoT platform. Their attributes include \texttt{name}, credentials consisting of a \texttt{username} and a \texttt{password}, and a \texttt{role}, which is either admin or user. Users can have multiple devices associated with their account and each device can have multiple sensors associated with it. Consumers are responsible for retrieving sensor data from the IoT platform. A user can have multiple consumers, that each allow access to a self-administered set of the user's sensors. Prior to retrieving data, the user has to grant a consumer access to one or multiple sensors (\S\ref{sec:sysdesign}). The relationship between the different entities and their attributes is shown in Figure~\ref{fig:iotplatformdata}.

\subsection{System Design}
\label{sec:sysdesign}
\begin{figure}[t]
\centering
\includegraphics[width=\columnwidth]{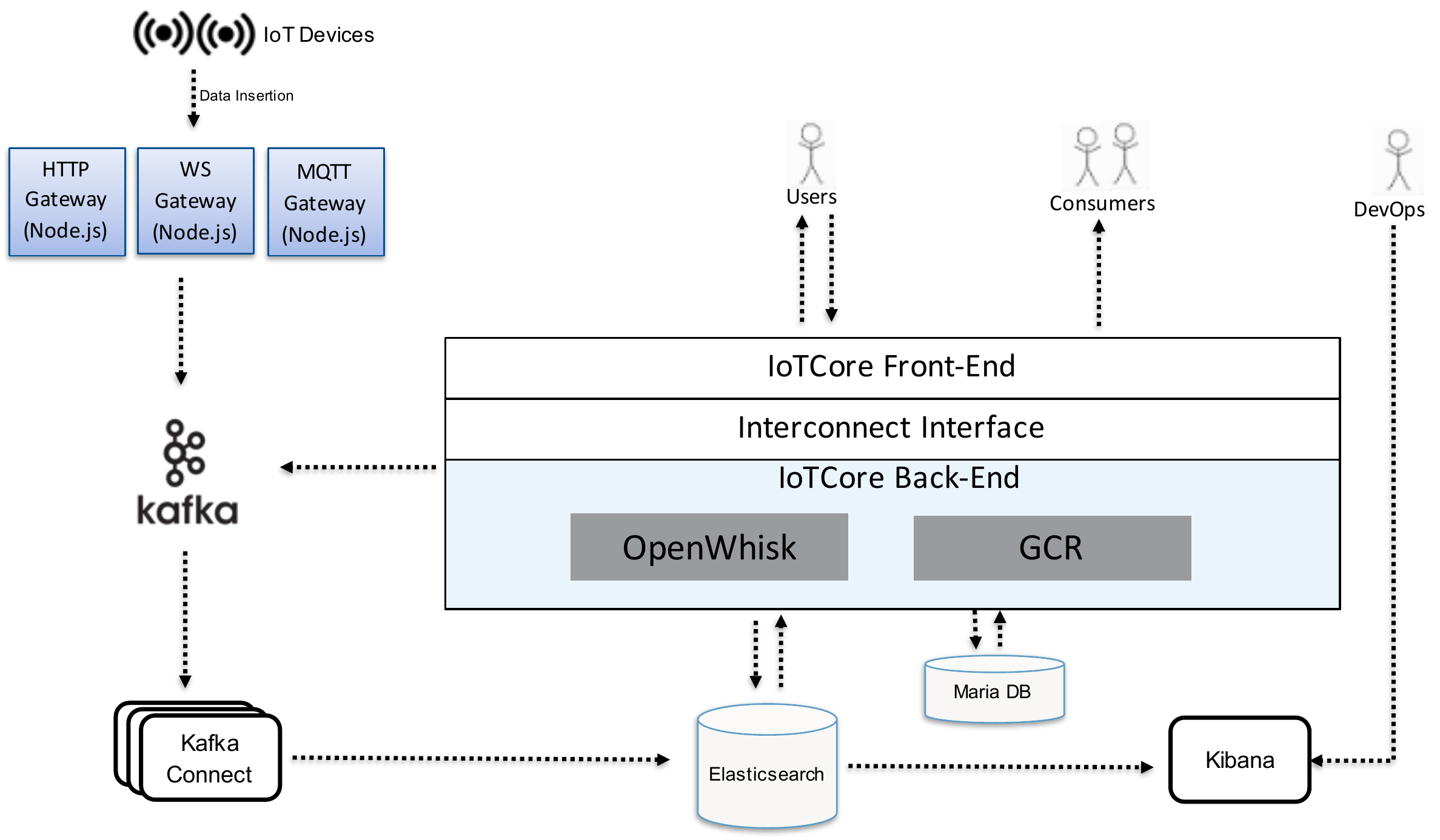}
\caption{Different components of the IoT platform.} 
\label{fig:iotplatform}
\shrinkspace
\end{figure}

The IoT platform consists of several independent microservices that interact with each other to provide the different functionalities described in \S\ref{sec:iotplatoverview} and \S\ref{sec:datamodel}. The different components of the platform are shown in Figure~\ref{fig:iotplatform}.

At the center of the platform is the \texttt{IoTCore} which is implemented as a web application with a \texttt{React} based front-end and a \texttt{Node.js} based backend. It enables users to manage devices, sensors, and authentication/authorization tokens for secure communications with the end devices. All data related to the users, devices, and sensors is stored in the relational database management system \texttt{MariaDB}. The end devices can send data to the platform using three different gateways, i.e.,  \texttt{HTTP}, \texttt{WebSocket} or \texttt{MQTT}. The main purpose of the gateway is to check the sensor's authentication token and verify its data schema. Currently our platform supports sending and storing data in various formats such as string, integers, or as floating point values. Moreover, it supports complex data types such as tuples and arrays. Following a successful verification by the gateway, the received sensor data is forwarded to \texttt{Kafka}. Kafka is a scalable streaming platform and incorporates a publish-subscribe pattern. One or multiple producers publish data records to a \textit{topic}, the core abstraction of Kafka that is used to store and process data record streams. A single \textit{topic} can then be subscribed by zero, one or multiple consumers (subscribers). Each sensor is matched to a separate Kafka topic in our current implementation. As a result, the received sensor data is logged under its own Kafka topic. Furthermore, for scalability purposes, each Kafka topic has a corresponding consumer called \texttt{Kafka-connect} for consuming  the data from the Kafka topic assigned to it and saving it into \texttt{Elasticsearch} (ES).  \texttt{Kafka-connect} enables the streaming of data between Kafka and other systems such as databases and cloud services. In the IoT platform it utilizes the ES service sink connector to write data from a topic in Kafka to an index in ES. ES is an open-source scalable search and analytics engine based on Apache Lucene. After receiving the sensor data in ES it is parsed, normalized and enriched before being indexed. An ES index stores data in a data structure called \textit{inverted index} that allows near real-time searches. All sensor data is persistently stored in ES. In our current implementation, one ES index is created for each sensor. The stored sensor data can be visualized using Kibana or can be retrieved by creating \textit{consumers} (\S\ref{sec:datamodel}).

\subsection{Interacting with the IoT Platform}
\label{sec:interaction}
The IoT platform provides several RESTful API endpoints that enable \textit{users} and \textit{consumers} to interact with the platform and perform create, read, update, and delete (CRUD) operations on the \textit{devices} and the \textit{sensors}. In a typical usecase, an admin user of the IoT platform creates new users by sending a post request to the \texttt{Users-Add} endpoint along with their attributes (\S\ref{sec:datamodel}). On receiving the request, new users are created with their attributes which are stored in MariaDB. Following this, users with an account on the IoT platform can login to the platform by sending a post request to the \texttt{Users-Signin} endpoint containing their username and password as parameters. Internally these credentials are compared to the values stored in MariaDB. After successful verification, a \texttt{JWT} is returned as the response which encodes the user's id. Following this, users can add their devices to the platform by sending a post request to the \texttt{Devices-Add} endpoint along with its attributes (\S\ref{sec:datamodel}) and the returned \texttt{JWT} in the authorization header. On successful addition to the database the added device id is returned as a response to the user. After this, users can create sensors for their devices by sending a post request to the \texttt{Sensors-Add} endpoint along with its attributes, the returned \texttt{JWT}, and the device id. On receiving the post request, the platform creates a Kafka topic and an ES index for the sensor. Moreover, it creates a Kafka-connect consumer for writing data from the Kafka topic to the ES index (\S\ref{sec:sysdesign}). On successful creation, the added id of the sensor is returned as a response to the user. Users can also obtain the different sensors associated with a particular device by sending a get request to the \texttt{Sensors-Get} endpoint along with the \texttt{JWT} and the device id. After adding a sensor, users can send sensor data to the platform via a post request to any of the different gateways (\S\ref{sec:sysdesign}). The post request should contain user id, device id, sensor id, payload, and a \texttt{JWT} authentication token associated with the device. The token can be obtained by sending a get request to the \texttt{DeviceKey-Get} endpoint along with the user id and the device id. For obtaining the sensor data from the platform, the user first needs to create a consumer by sending a post request with its user id and \texttt{JWT} to the \texttt{Consumer-Add} endpoint. Following this, the user needs to grant the created consumer access to the sensor by sending another post request with its user id, consumer id, sensor id, and \texttt{JWT} to the \texttt{ConsumerSensor-Enable} endpoint. Finally, the sensor data can be obtained by sending a get request to the \texttt{Consumers-Consume-Get} endpoint along with the sensor id and the \texttt{JWT} token associated with the consumer. Similar to devices the token for a consumer can be obtained by sending a get request to the \texttt{ConsumerKey-Get} endpoint. On successful verification of the request parameters, the sensor data is retrieved from ES and sent as a response to the user. Overall the platform provides 30 different API endpoints all of which are implemented in the \texttt{IoTCore} backend. Note that a user can also use the \texttt{IoTCore} front-end for performing the different CRUD operations.



\section{Migration Methodology}
\label{sec:methodology}
In this work, we migrate the IoT platform application onto Apache OpenWhisk (OW) \cite{openwhisk} and Google Cloud Run (GCR) \cite{GCR}. Our application consists of several \textit{off-the-shelf} software components, i.e., \texttt{Kafka}, \texttt{Kafka-connect}, \texttt{MariaDB}, \texttt{Elastic Search}, and \texttt{Kibana} (\S\ref{sec:sysdesign}). To minimize changes to the application design, we do not migrate these components. As a result, we only focus on adapting and decomposing the \texttt{IoTCore} backend component to utilize serverless functions.

The \texttt{IoTCore} backend component uses the \texttt{Express} framework to serve incoming \textit{user} and \textit{consumer} requests (\S\ref{sec:sysdesign}). A router checks on which API endpoint a request was received and then forwards it to the designated middleware functions responsible for interacting with the different \textit{off-the-shelf} software components. As part of our migration of the \texttt{IoTCore} backend, we decompose the application logic for each specific API endpoint into a separate function. We describe the migration process for one specific API endpoint, namely, \texttt{Sensors-Get}, while the other API endpoints were adapted similarly. This API endpoint is responsible for obtaining the different sensors associated with a particular device. For the \texttt{Sensors-Get} API endpoint, an incoming request is first forwarded to the \textit{authentication handler} and then to the \textit{sensors} controller which obtains the number of sensors attached to the device from \texttt{MariaDB} and returns them to the user. To migrate the application logic for this endpoint to OW these two middleware components were extracted into individual functions. Towards this, we created custom function runtimes based on the \texttt{Node.js-action} Docker images~\cite{nodejsaction}. The runtimes contain only the required \texttt{Node.js} packages required for the two middleware components along with connection configurations and \texttt{SQL} schema. As a result, the docker images for the individual functions are smaller in size as compared to the docker image of the \texttt{IoTCore} which contains every package required to respond to requests for all API endpoints. During our migration to OW, we could reuse the \texttt{IoTCore} backend's original code to a large extent since the individual functions of the different controllers, i.e., \textit{users}, \textit{devices}, \textit{sensors}, and \textit{consumers} closely mirrored the requirements of single OW functions. Moreover, we utilized the function sequencing functionality provided by OW to chain the \textit{authentication} function with the different middleware functions. Since authentication is required for almost all API endpoints it was possible to reuse already running \textit{authentication} function instances for requests at different endpoints. For migration onto GCR, we followed a similar approach. With GCR the application code reusage was even higher since it utilizes the \texttt{Express} framework~\cite{google2022b} used in our application. However, it does not provide the functionality of chaining functions together like in OW. Therefore, we added the application logic for authentication and API endpoint within the same function. While it is possible to implement function sequencing using Google Cloud Workflows, it is a different service which is billed separately and is out of scope for this work.

We migrated the functionality of all the different API endpoints provided by the \texttt{IoTCore} to OW and GCR (\S\ref{sec:sysdesign}). In addition, we implemented an \texttt{Interconnect} interface as shown in Figure~\ref{fig:iotplatform}. This interface intercepts an incoming request to an API endpoint and forwards it to the appropriate OW or GCR function. Moreover, it is responsible for forwarding the function response to the user after it completes execution. Finally, we also migrated the \texttt{HTTP} gateway into separate OW and GCR functions (\S\ref{sec:sysdesign}). This was possible since it is stateless and uses RESTful APIs for communication . In contrast, \texttt{WebSocket} and \texttt{MQTT} protocols are inherently stateful which prevented their adaption into OW and GCR functions~\cite{ws, aedes}.

\begin{figure*}
 \begin{subfigure}{0.33\textwidth}
    \centering
        \includegraphics[width=\columnwidth]{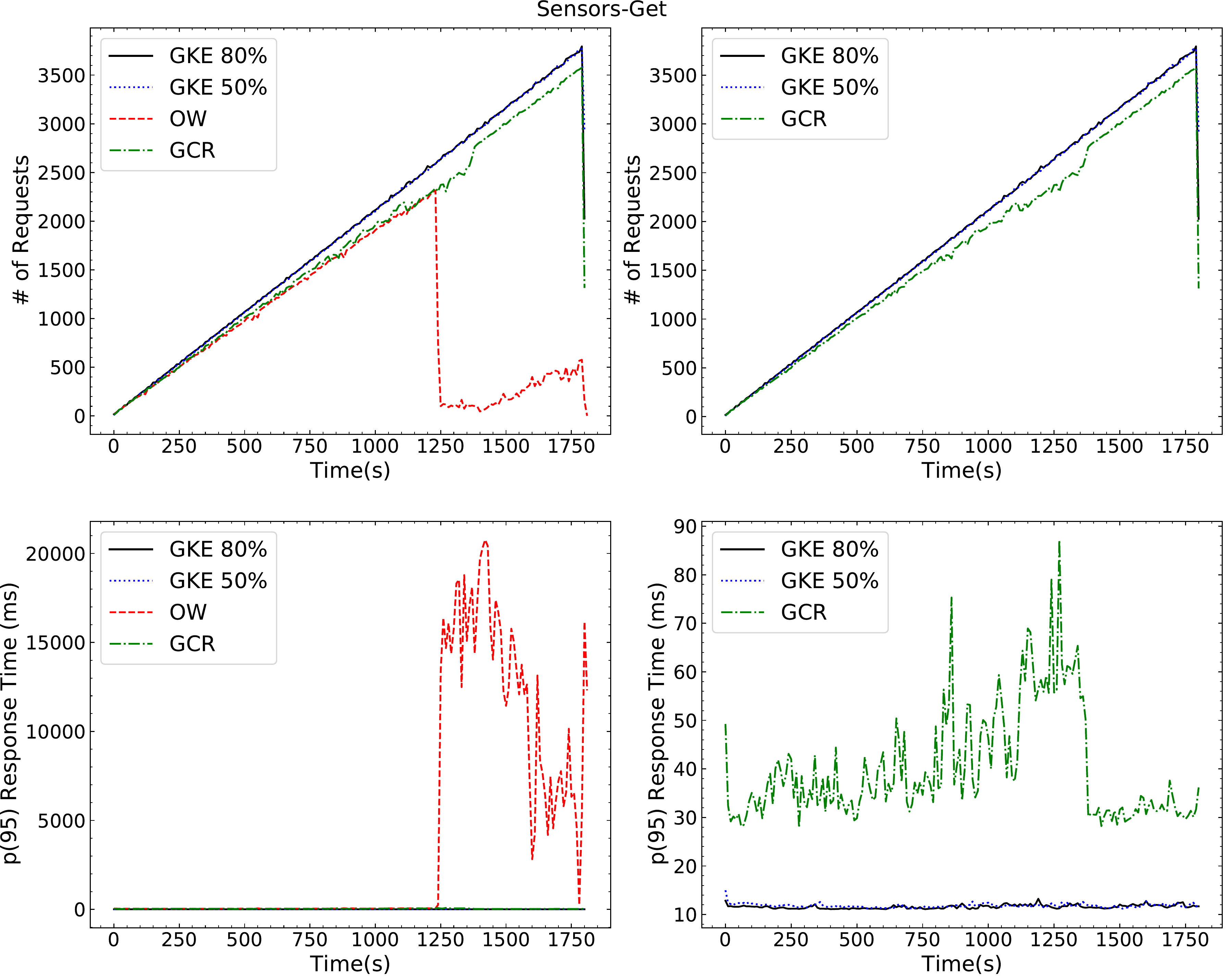}
        \caption{Linear scenario.}
        \label{fig:linearsensorsget}
    \end{subfigure}
\begin{subfigure}{0.33\textwidth}
    \centering
        \includegraphics[width=\columnwidth]{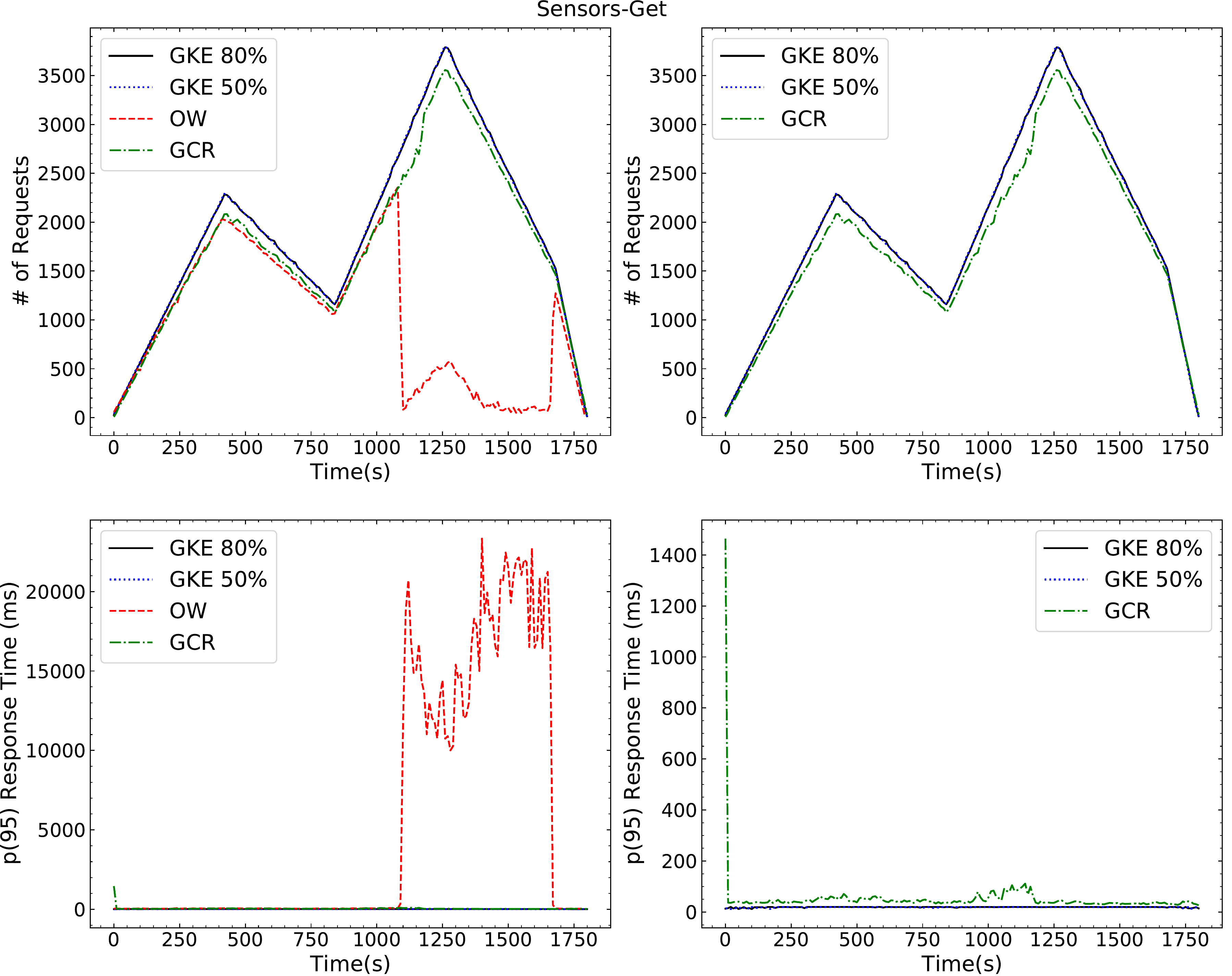}
        \caption{Random scenario.}
        \label{fig:randomsensorsget}
\end{subfigure}
\begin{subfigure}{0.33\textwidth}
    \centering
         \includegraphics[width=\columnwidth]{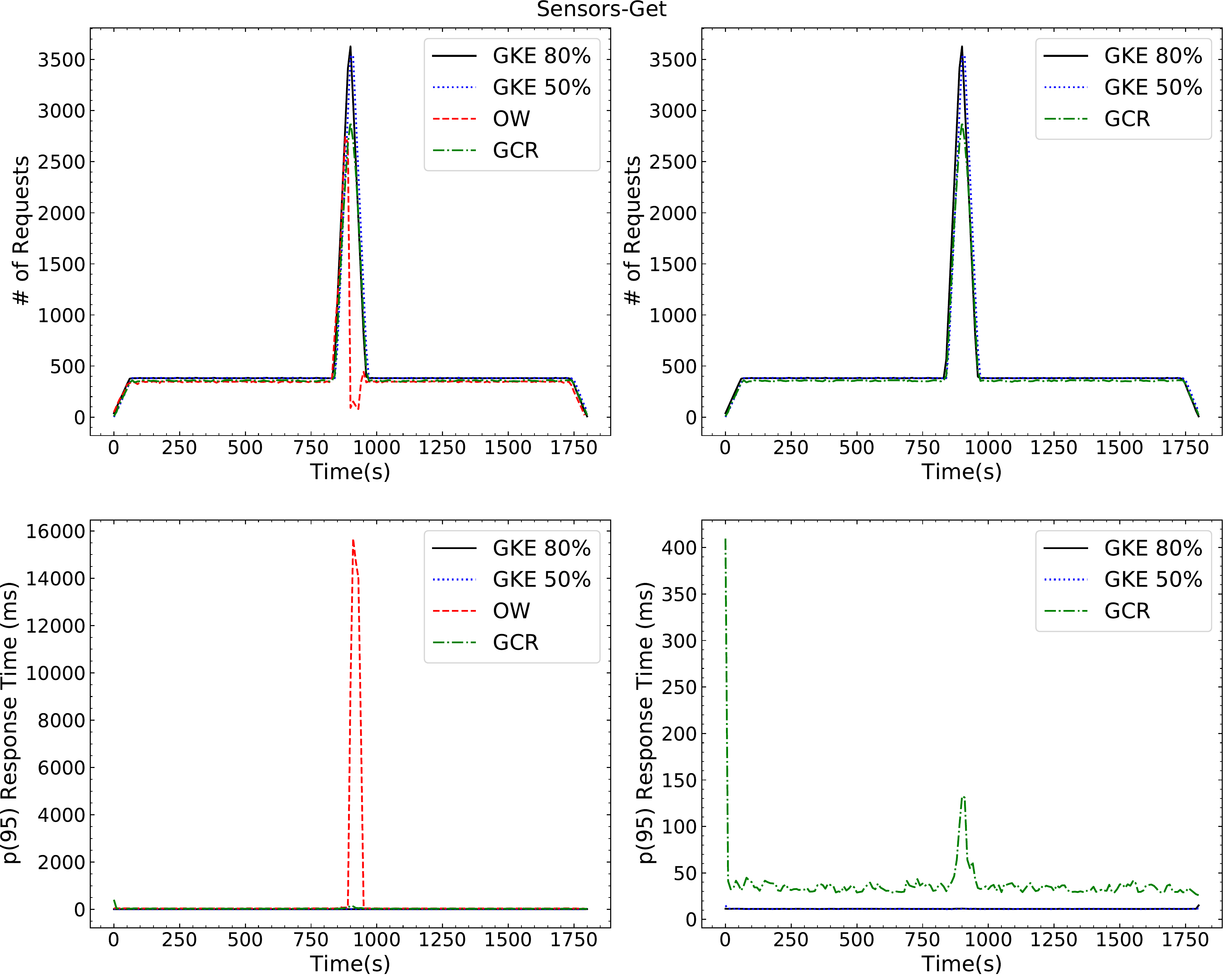}
        \caption{Spike scenario.}
        \label{fig:spikesensorsget}
\end{subfigure}
 \begin{subfigure}{0.33\textwidth}
    \centering
        \includegraphics[width=\columnwidth]{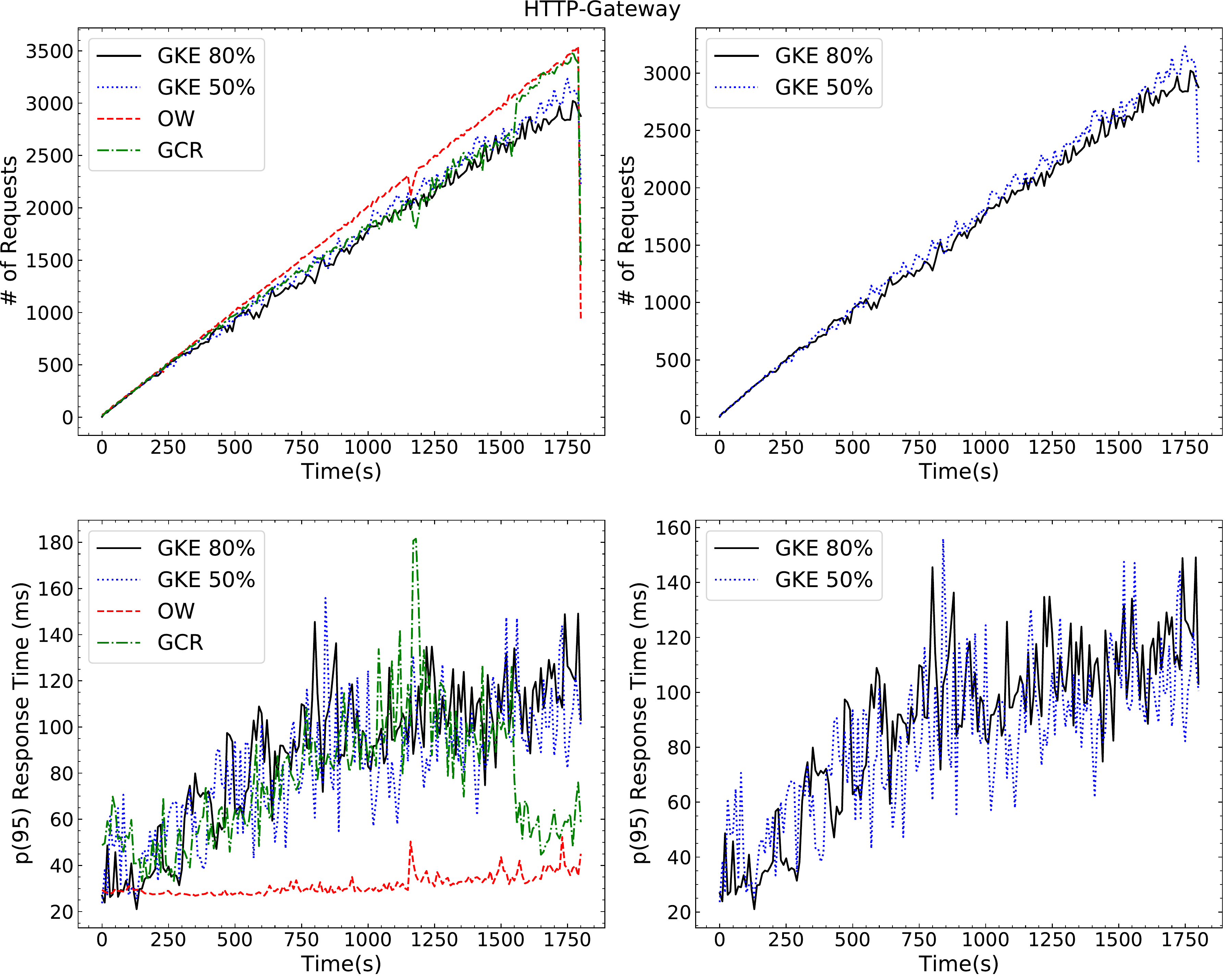}
        \caption{Linear scenario.}
        \label{fig:linearhttpgateway}
    \end{subfigure}
\begin{subfigure}{0.33\textwidth}
    \centering
        \includegraphics[width=\columnwidth]{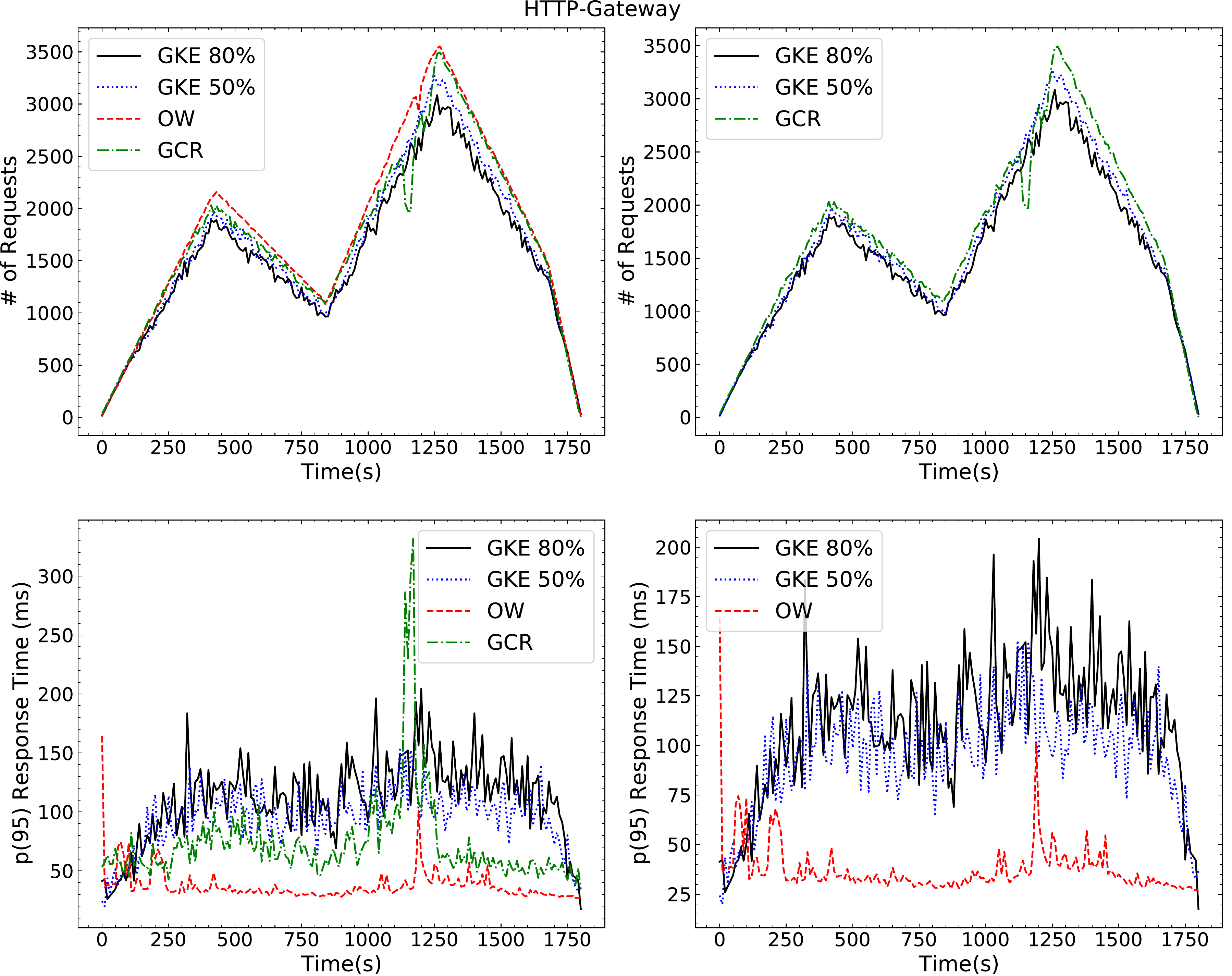}
        \caption{Random scenario.}
        \label{fig:randomhttpgateway}
\end{subfigure}
\begin{subfigure}{0.33\textwidth}
    \centering
         \includegraphics[width=\columnwidth]{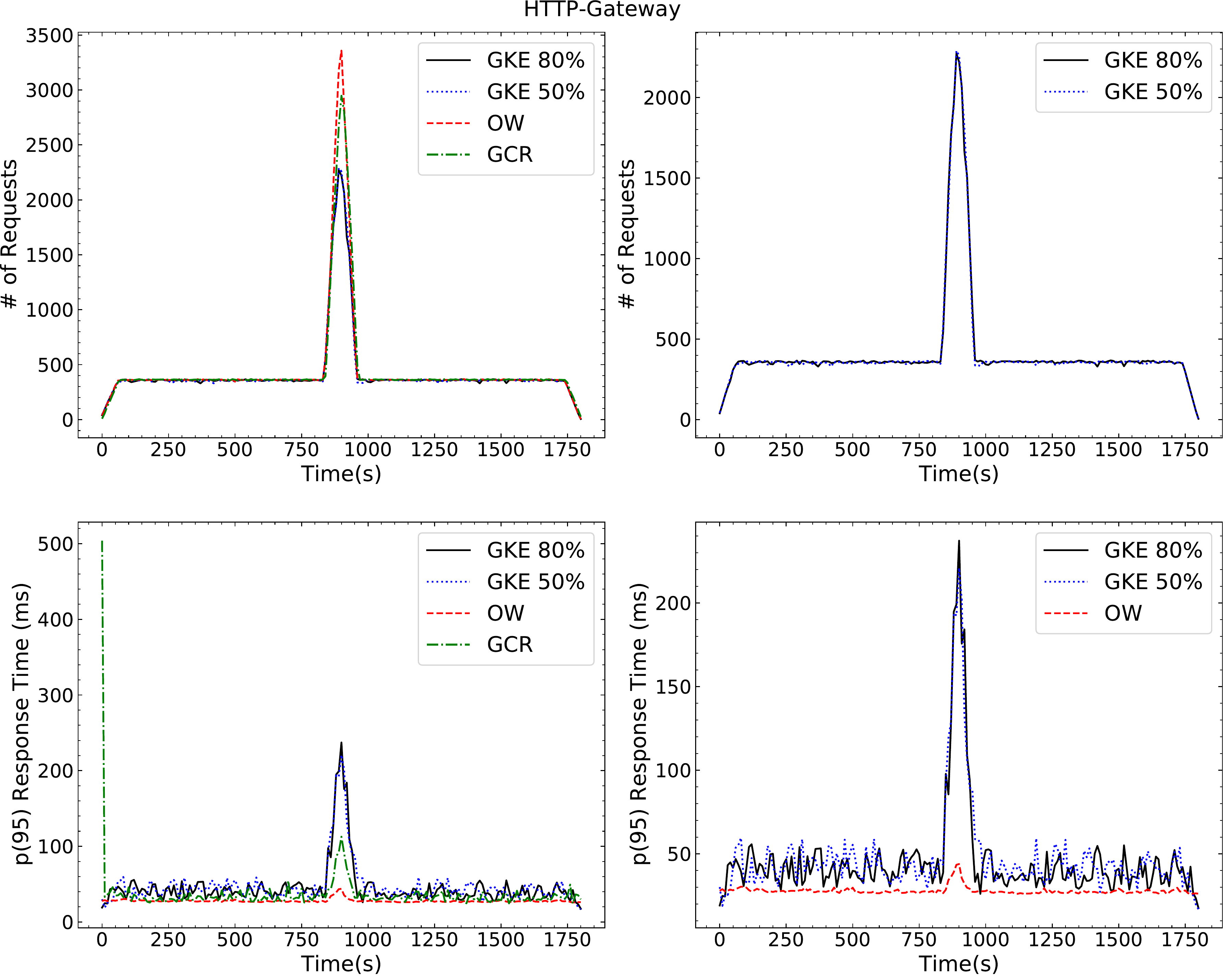}
        \caption{Spike scenario.}
        \label{fig:spikehttpgateway}
\end{subfigure}

\caption{Performance comparison for the \texttt{Sensor-Get} endpoint and \texttt{HTTP} gateway across the different deployment strategies (\S\ref{sec:deploymentconfig}) and load testing scenarios (\S\ref{sec:testingscenarios}). All metric values are sampled on 10 seconds intervals.}
\label{fig:perfresults}
\shrinkspace
\end{figure*}

\section{Experimental Results}
\label{sec:results}
In this section, we present results wrt performance and cost for the microservices and serverless versions of the IoT platform. In addition, we highlight the lessons learned from our migration. To limit the total number of experiments, we only focus on six relevant API endpoints (\S\ref{sec:perfcomp},\S\ref{sec:costcomp}). For all our experiments, we follow best practices while reporting results.


\subsection{Experiment Setup}
\label{sec:expconfig}
To compare the performance of the original microservice architecture of our IoT platform with our migrated versions (\S\ref{sec:methodology}), we consider three deployment strategies: (i) Google Kubernetes Engine (GKE) in standard mode~\cite{gke}, (ii) OW on top of GKE, and (iii) GCR. 


\subsubsection{Deployment Configuration}
\label{sec:deploymentconfig}
To guarantee fairness between the microservice and serverless deployment strategies, we deploy all \textit{off-the-shelf} software components (\S\ref{sec:sysdesign}) to a separate Kubernetes cluster with a fixed resource configuration of three VMs of type \texttt{e2-standard-4}~\cite{gcpvms}, i.e., 4vCPUs and 16 GiB of memory. This cluster contains all components except the \texttt{IoTCore} and the \texttt{HTTP} gateway. All services were exposed using internal load balancers that make them accessible only within the project's virtual private cloud (VPC). This also guarantees that access times were comparable between all three deployments strategies.

Following this, for the first deployment strategy, we deploy the services under investigation on a GKE cluster with a node pool of two VMs of type \texttt{e2-standard-8}~\cite{gcpvms} and 100 GiB of persistent disk memory. We set the maximum number of nodes for the GKE cluster to five and enable on-demand \textit{cluster-autoscaling}~\cite{google2022f}. We configure the \texttt{IoTCore} and the \texttt{HTTP} gateway deployments to request $0.5$vCPU and use a maximum of one vCPU per pod. Furthermore, we enable horizontal pod autoscaling (HPA) for the GKE cluster~\cite{kubernetes2022b} with a maximum limit of $40$ pods for the two different services. For our experiments, we consider two different HPA configurations, i.e., $50$\% CPU utilization (GKE-50) and $80$\% CPU utilization (GKE-80). For our OW setup, we use a GKE cluster with a node pool of five VMs of type  \texttt{e2-standard-8}. We set the initial number of nodes to five to keep the performance results comparable with our first deployment strategy. The individual OW functions were configured with a maximum memory of $256$ MiB (\S\ref{sec:methodology}). Moreover, we modify the default configurations of the different components of OW to enhance scalability~\cite{openwhisk2021b}. Towards this, we set the number of \textit{controllers} and \textit{invokers} to two, enable request concurrency, increase the user pool memory to $10$GiB, disable logging, and set the heap space for the \textit{controller} and \textit{invoker} to $2$GiB and $1$GiB respectively. In addition, we limit the number of action invocations to $25000$ per minute and the maximum concurrent invocations to $9999$. For the third deployment strategy, we configure the individual GCR functions to use one vCPU with $256$ MiB of memory. We set the maximum number of instances for a GCR function to 20 and limit the request concurrency to 100. Moreover, we connect the GCR functions to the shared services with a serverless VPC connector~\cite{google2022g}, that enables access to the internal load balancers of the \textit{off-the-shelf} services.


\subsubsection{Load Testing Scenarios}
\label{sec:testingscenarios}
For generating load at the different API endpoints (\S\ref{sec:sysdesign}), we use the open-source performance and regression testing tool \texttt{k6}. It uses a script for running the load tests where the API endpoint along with the request parameters are specified. Tests in  \texttt{k6} are based on virtual users (VUs), which are entities that make HTTP(s) requests and try to perform a given test as often as possible. The number of requests per second (RPS) generated by \texttt{k6} depends on the number of VUs and the time taken by each request to complete. For our experiments, we consider three different scenarios that mirror different workloads. First, \textbf{linear}, i.e., a 30 minutes (mins) linear increase to a target of 100 VUs. Second, \textbf{random}, i.e.,  60 VUs at 7 mins, 30 VUs at 14 mins, 100 VUs at 21 mins, 40 VUs at 28 mins, and zero VUs after 30 mins. The transitions between targeted values are linear. Third, \textbf{spike}, i.e., a plateau of 10 VUs is reached after one min. Following this, a spike of 100 VUs is created between 14 and 16 mins. After the spike, the plateau of 10 VUs is held for another 13 mins before the VUs are linearly decreased to zero. All scenarios are executed for 30 mins.


For our experiments on the different API endpoints, we create dummy users and data on the IoT platform. For each request, a random user with the appropriate parameter is selected. All load tests are executed on a VM instance hosted on the LRZ compute cloud~\cite{lrzcc} with a configuration of 10vCPUs and 45 GiB of RAM. We ensure sufficient time between each test to allow the different deployment strategies to scale back to their initial state. 





\subsection{Comparing Performance}
\label{sec:perfcomp}
To compare the performance across the different deployment strategies (\S\ref{sec:deploymentconfig}) and load testing scenarios (\S\ref{sec:testingscenarios}), we consider metrics based on request response times, i.e., average and p(95), and the number of successful requests, i.e., requests with a status code of $200$. The performance results for the \texttt{Sensor-Get} API endpoint for the different load testing scenarios are shown in Figures~\ref{fig:linearsensorsget},~\ref{fig:randomsensorsget}, and~\ref{fig:spikesensorsget}. For the linear workload, we observe that GKE-80 was able to serve $343,343$ total requests with an aggregated p(95) response time of $19.63$ milliseconds (ms), while GKE-50 served $343,158$ requests with a p(95) response time of $19.72$ms. On the other hand, GCR served $319,046$ requests within $53.74$ms and OW served $161,053$ requests within $94.15$ms. We observe a significant increase in the p(95) response time for OW for greater than $2000$ requests per ten seconds. This can be attributed to the overhead for executing \textit{sequence actions} in OW (\S\ref{sec:methodology}) for a large number of concurrent requests. Moreover, we observe that all requests were successfully handled in GKE-50, GKE-80, and GCR deployment strategies, while with OW eight requests failed. For the random workload, GKE-50 and GKE-80 served $346,864$ and $346,620$ requests within $19.54$ms and $19.65$ms aggregated p(95) response times respectively. GCR served $321,963$ requests within $52.86$ms, while OW served only $175,065$ requests within $86.75$ms. Similarly, for the spike workload GKE-50 served a total of $87,132$ requests, GKE-80 $87,085$ requests, GCR $79,628$ requests, and OW $68,409$ requests. The aggregated p(95) response times for the different strategies were $19.59$ms, $19.76$ms, $65.11$ms and $69.73$ms respectively. For both random and spike workloads GKE-50, GKE-80, and GCR were able to serve all requests successfully, while with OW nine and four requests failed. The initial peaks in the p(95) response times for GCR can be attributed to \textit{cold starts}.

\begin{table}[t]
\centering

\begin{adjustbox}{width=8.7cm,  center}
\begin{tabular}{|c|c|ccccccccc|}
\hline
\multirow{3}{*}{API Endpoint}         & \multicolumn{1}{l|}{\multirow{3}{*}{Deployment Strategy}} & \multicolumn{9}{c|}{Load Testing Scenarios}                                                                                                                                                                                                                                                                                         \\ \cline{3-11} 
                                      & \multicolumn{1}{l|}{}                                     & \multicolumn{3}{c|}{Linear}                                                                                 & \multicolumn{3}{c|}{Random}                                                                                 & \multicolumn{3}{c|}{Spike}                                                                              \\ \cline{3-11} 
                                      & \multicolumn{1}{l|}{}                                     & \multicolumn{1}{c|}{\#Requests} & \multicolumn{1}{c|}{Average (ms)}   & \multicolumn{1}{c|}{p(95) (ms)}     & \multicolumn{1}{r|}{\#Requests} & \multicolumn{1}{l|}{Average (ms)}   & \multicolumn{1}{l|}{p(95) (ms)}     & \multicolumn{1}{l|}{\#Requests} & \multicolumn{1}{l|}{Average (ms)}   & \multicolumn{1}{l|}{p(95) (ms)} \\ \hline
\multirow{4}{*}{Sensors-Get}          & GKE-50                                                    & \multicolumn{1}{c|}{343,158}    & \multicolumn{1}{c|}{11.47}          & \multicolumn{1}{c|}{19.72}          & \multicolumn{1}{c|}{346,864}    & \multicolumn{1}{c|}{\textbf{11.18}} & \multicolumn{1}{c|}{\textbf{19.54}} & \multicolumn{1}{c|}{87,132}     & \multicolumn{1}{c|}{\textbf{11.24}} & \textbf{19.59}                  \\ \cline{2-11} 
                                      & GKE-80                                                    & \multicolumn{1}{c|}{343,343}    & \multicolumn{1}{c|}{\textbf{11.35}} & \multicolumn{1}{c|}{\textbf{19.63}} & \multicolumn{1}{c|}{346,620}    & \multicolumn{1}{c|}{11.39}          & \multicolumn{1}{c|}{19.65}          & \multicolumn{1}{c|}{87,085}     & \multicolumn{1}{c|}{11.38}          & 19.76                           \\ \cline{2-11} 
                                      & OW                                                        & \multicolumn{1}{c|}{161,053}    & \multicolumn{1}{c|}{310.08}         & \multicolumn{1}{c|}{94.15}          & \multicolumn{1}{c|}{176,065}    & \multicolumn{1}{c|}{267.47}         & \multicolumn{1}{c|}{86.75}          & \multicolumn{1}{c|}{68,409}     & \multicolumn{1}{c|}{88.80}          & 65.11                           \\ \cline{2-11} 
                                      & GCR                                                       & \multicolumn{1}{c|}{319,046}    & \multicolumn{1}{c|}{31.29}          & \multicolumn{1}{c|}{53.74}          & \multicolumn{1}{c|}{321,963}    & \multicolumn{1}{c|}{31.45}          & \multicolumn{1}{c|}{52.86}          & \multicolumn{1}{c|}{79,628}     & \multicolumn{1}{c|}{36.05}          & 69.73                           \\ \hline
\multirow{4}{*}{HTTP-Gateway}         & GKE-50                                                    & \multicolumn{1}{c|}{295,154}    & \multicolumn{1}{c|}{54.16}          & \multicolumn{1}{c|}{113.41}         & \multicolumn{1}{c|}{297,088}    & \multicolumn{1}{c|}{55.11}          & \multicolumn{1}{c|}{109.61}         & \multicolumn{1}{c|}{75,969}     & \multicolumn{1}{c|}{49.87}          & 148.34                          \\ \cline{2-11} 
                                      & GKE-80                                                    & \multicolumn{1}{c|}{283,433}    & \multicolumn{1}{c|}{66.82}          & \multicolumn{1}{c|}{123.64}         & \multicolumn{1}{c|}{283,071}    & \multicolumn{1}{c|}{70.29}          & \multicolumn{1}{c|}{130.61}         & \multicolumn{1}{c|}{76,256}     & \multicolumn{1}{c|}{48.76}          & 152.55                          \\ \cline{2-11} 
                                      & OW                                                        & \multicolumn{1}{c|}{322,721}    & \multicolumn{1}{c|}{\textbf{28.02}} & \multicolumn{1}{c|}{\textbf{39.33}} & \multicolumn{1}{c|}{326,900}    & \multicolumn{1}{c|}{\textbf{27.15}} & \multicolumn{1}{c|}{\textbf{37.63}} & \multicolumn{1}{c|}{81,971}     & \multicolumn{1}{c|}{\textbf{27.74}} & \textbf{33.61}                  \\ \cline{2-11} 
                                      & GCR                                                       & \multicolumn{1}{c|}{299,420}    & \multicolumn{1}{c|}{49.73}          & \multicolumn{1}{c|}{108.39}         & \multicolumn{1}{c|}{311,514}    & \multicolumn{1}{c|}{40.88}          & \multicolumn{1}{c|}{88.38}          & \multicolumn{1}{c|}{81,029}     & \multicolumn{1}{c|}{30.98}          & 72.45                           \\ \hline
\multirow{4}{*}{Users-Get}            & GKE-50                                                    & \multicolumn{1}{c|}{327,921}    & \multicolumn{1}{c|}{\textbf{23.70}} & \multicolumn{1}{c|}{\textbf{36.50}} & \multicolumn{1}{c|}{329,417}    & \multicolumn{1}{c|}{25.09}          & \multicolumn{1}{c|}{42.25}          & \multicolumn{1}{c|}{81,725}     & \multicolumn{1}{c|}{28.69}          & 62.50                           \\ \cline{2-11} 
                                      & GKE-80                                                    & \multicolumn{1}{c|}{324,931}    & \multicolumn{1}{c|}{26.20}          & \multicolumn{1}{c|}{47.90}          & \multicolumn{1}{c|}{328,549}    & \multicolumn{1}{c|}{25.82}          & \multicolumn{1}{c|}{44.07}          & \multicolumn{1}{c|}{82,791}     & \multicolumn{1}{c|}{25.09}          & 43.56                           \\ \cline{2-11} 
                                      & OW                                                        & \multicolumn{1}{c|}{154,893}    & \multicolumn{1}{c|}{332.29}         & \multicolumn{1}{c|}{145.88}         & \multicolumn{1}{c|}{259,701}    & \multicolumn{1}{c|}{590.08}         & \multicolumn{1}{c|}{2290}           & \multicolumn{1}{c|}{61,472}     & \multicolumn{1}{c|}{127.90}         & 52.44                           \\ \cline{2-11} 
                                      & GCR                                                       & \multicolumn{1}{c|}{325,239}    & \multicolumn{1}{c|}{25.94}          & \multicolumn{1}{c|}{38.36}          & \multicolumn{1}{c|}{330,382}    & \multicolumn{1}{c|}{\textbf{24.28}} & \multicolumn{1}{c|}{\textbf{35.58}} & \multicolumn{1}{c|}{83,120}     & \multicolumn{1}{c|}{\textbf{23.94}} & \textbf{32.75}                  \\ \hline
\multirow{4}{*}{Devices-Add}          & GKE-50                                                    & \multicolumn{1}{c|}{298,950}    & \multicolumn{1}{c|}{50.29}          & \multicolumn{1}{c|}{93.35}          & \multicolumn{1}{c|}{307,056}    & \multicolumn{1}{c|}{45.20}          & \multicolumn{1}{c|}{76.89}          & \multicolumn{1}{c|}{80,252}     & \multicolumn{1}{c|}{\textbf{33.83}} & 89.10                           \\ \cline{2-11} 
                                      & GKE-80                                                    & \multicolumn{1}{c|}{280,226}    & \multicolumn{1}{c|}{70.41}          & \multicolumn{1}{c|}{125.90}         & \multicolumn{1}{c|}{279,296}    & \multicolumn{1}{c|}{74.61}          & \multicolumn{1}{c|}{135.03}         & \multicolumn{1}{c|}{78,108}     & \multicolumn{1}{c|}{41.68}          & 124.49                          \\ \cline{2-11} 
                                      & OW                                                        & \multicolumn{1}{c|}{176,729}    & \multicolumn{1}{c|}{261.33}         & \multicolumn{1}{c|}{\textbf{66.40}} & \multicolumn{1}{c|}{166,670}    & \multicolumn{1}{c|}{296.90}         & \multicolumn{1}{c|}{78.76}          & \multicolumn{1}{c|}{65,377}     & \multicolumn{1}{c|}{105.43}         & \textbf{52.15}                  \\ \cline{2-11} 
                                      & GCR                                                       & \multicolumn{1}{c|}{310,090}    & \multicolumn{1}{c|}{\textbf{39.45}} & \multicolumn{1}{c|}{74.11}          & \multicolumn{1}{c|}{318,868}    & \multicolumn{1}{c|}{\textbf{34.20}} & \multicolumn{1}{c|}{\textbf{62.10}} & \multicolumn{1}{c|}{79,616}     & \multicolumn{1}{c|}{36.05}          & 72.29                           \\ \hline
\multirow{4}{*}{Devices-Get}          & GKE-50                                                    & \multicolumn{1}{c|}{334,605}    & \multicolumn{1}{c|}{\textbf{18.18}} & \multicolumn{1}{c|}{\textbf{26.35}} & \multicolumn{1}{c|}{337,044}    & \multicolumn{1}{c|}{\textbf{18.81}} & \multicolumn{1}{c|}{\textbf{24.49}} & \multicolumn{1}{c|}{80,429}     & \multicolumn{1}{c|}{33.17}          & 131.56                          \\ \cline{2-11} 
                                      & GKE-80                                                    & \multicolumn{1}{c|}{332,665}    & \multicolumn{1}{c|}{19.75}          & \multicolumn{1}{c|}{29.13}          & \multicolumn{1}{c|}{335,255}    & \multicolumn{1}{c|}{20.26}          & \multicolumn{1}{c|}{28.40}          & \multicolumn{1}{c|}{83,207}     & \multicolumn{1}{c|}{\textbf{23.64}} & 62.57                           \\ \cline{2-11} 
                                      & OW                                                        & \multicolumn{1}{c|}{227,558}    & \multicolumn{1}{c|}{146.35}         & \multicolumn{1}{c|}{76.99}          & \multicolumn{1}{c|}{162,407}    & \multicolumn{1}{c|}{310.86}         & \multicolumn{1}{c|}{101.04}         & \multicolumn{1}{c|}{62,137}     & \multicolumn{1}{c|}{125.25}         & \textbf{52.14}                  \\ \cline{2-11} 
                                      & GCR                                                       & \multicolumn{1}{c|}{322,234}    & \multicolumn{1}{c|}{28.48}          & \multicolumn{1}{c|}{39.29}          & \multicolumn{1}{c|}{324,856}    & \multicolumn{1}{c|}{28.93}          & \multicolumn{1}{c|}{37.55}          & \multicolumn{1}{c|}{79,616}     & \multicolumn{1}{c|}{36.05}          & 72.29                           \\ \hline
\multirow{4}{*}{Consumer-Consume-Get} & GKE-50                                                    & \multicolumn{1}{c|}{312,530}    & \multicolumn{1}{c|}{37.18}          & \multicolumn{1}{c|}{83.18}          & \multicolumn{1}{c|}{318,553}    & \multicolumn{1}{c|}{\textbf{34.48}} & \multicolumn{1}{c|}{72.17}          & \multicolumn{1}{c|}{81,593}     & \multicolumn{1}{c|}{\textbf{29.07}} & 61.77                           \\ \cline{2-11} 
                                      & GKE-80                                                    & \multicolumn{1}{c|}{311,265}    & \multicolumn{1}{c|}{38.37}          & \multicolumn{1}{c|}{86.08}          & \multicolumn{1}{c|}{316,838}    & \multicolumn{1}{c|}{36.02}          & \multicolumn{1}{c|}{77.74}          & \multicolumn{1}{c|}{81,478}     & \multicolumn{1}{c|}{29.45}          & 63.27                           \\ \cline{2-11} 
                                      & OW                                                        & \multicolumn{1}{c|}{314,545}    & \multicolumn{1}{c|}{\textbf{35.31}} & \multicolumn{1}{c|}{\textbf{52.23}} & \multicolumn{1}{c|}{317,326}    & \multicolumn{1}{c|}{35.57}          & \multicolumn{1}{c|}{\textbf{51.52}} & \multicolumn{1}{c|}{79,473}     & \multicolumn{1}{c|}{36.53}          & \textbf{45.68}                  \\ \cline{2-11} 
                                      & GCR                                                       & \multicolumn{1}{c|}{307,828}    & \multicolumn{1}{c|}{41.56}          & \multicolumn{1}{c|}{86.52}          & \multicolumn{1}{c|}{312,235}    & \multicolumn{1}{c|}{40.23}          & \multicolumn{1}{c|}{81.38}          & \multicolumn{1}{c|}{77,493}     & \multicolumn{1}{c|}{43.85}          & 115.31                          \\ \hline
\end{tabular}
\end{adjustbox}
\caption{Total number of requests, average and p(95) response times for the different API endpoints across the different deployment strategies and load testing scenarios. The highlighted values represent the minimum values of the response times for a particular workload.}
\label{tab:perfcomparison}
\shrinkspace
\end{table}


In contrast to the \texttt{Sensors-Get} API endpoint, we observe that with the \texttt{HTTP-Gateway}, serverless deployments, i.e., GCR and OW perform better than GKE-50 and GKE-80 as shown in Figures~\ref{fig:linearhttpgateway},~\ref{fig:randomhttpgateway}, and~\ref{fig:spikehttpgateway}. For the linear workload, OW served $322,721$ requests with an aggregated p(95) response time of $39.33$ms, while GCR served $299,420$ requests within $108.39$ms. GKE-50 and GKE-80 served $295,154$ and $283,433$ requests within $113.41$ms and $123.64$ms respectively. For the random workload, OW and GCR served $326,900$ and $311,514$ requests with aggregated p(95) response times of $37.63$ms and $88.38$ms respectively. On the other hand, GKE-50 and GKE-80 served $297,088$ and $283.071$ requests within $109.61$ms and $130.61$ms respectively. Similarly for the spike workload OW served $81,971$ requests within $33.61$ms, GCR served $81,029$ requests within $30.98$ms, GKE-50 served $75,969$ requests within $148.34$ms, and GKE-80 served $76,256$ requests within $152.55$ms. For all workloads GKE-50, GKE-80, and OW were able to serve all requests successfully, while with GCR nine and four requests failed for the linear and random workloads. The low p(95) response times observed for OW across the different workloads as compared to the different deployment strategies can be attributed to the high initial resource provisioning for our OW setup, i.e., five VMs (\S\ref{sec:deploymentconfig}). Moreover, while GCR limits the amount of CPU resources allocated to a function, with OW no such limit is enforced leading to better performance. Furthermore, implementation of the gateway function in OW does not require action sequencing.

Table~\ref{tab:perfcomparison} summarizes the performance results for the different API endpoints across the different deployment strategies and workloads. 
From our experiments, we observe that for the API endpoints \texttt{Sensors-Get}, \texttt{Users-Get}, and \texttt{Devices-Get}, the GKE standard deployment strategies outperform the serverless deployments. For the \texttt{HTTP} gateway OW performs best, while for the \texttt{Devices-Add} endpoint GCR delivers best performance. Although OW has lower response times for certain workloads than GCR for the \texttt{Devices-Add} endpoint, the number of requests served are significantly lower. Moreover, for the \texttt{Consumer-Consumer-Get} endpoint OW performs best for the linear and random workloads, while GKE-50 has the lowest average response time and maximum number of served requests for the spike workload. This endpoint is responsible for obtaining data from ES (\S\ref{sec:sysdesign}) and also does not require action sequencing.  Across the different load testing scenarios, we observed that GKE setups scaled well to handle the incoming requests. This behaviour was closely matched by GCR which delivered robust performance across the different scenarios and endpoints. However, with OW, we observed significantly higher response times at the peak of the different workload patterns leading to request failures (e.g. Figure~\ref{fig:linearsensorsget}). GKE-50 performs slightly better as compared to the GKE-80 deployment strategy. This can be attributed to the HPA configuration in GKE-50 which initiates execution of new pods earlier as compared GKE-80.





\begin{table}[t]
\centering
\begin{adjustbox}{width=8cm,  center}
\begin{tabular}{|c|c|clclcc|}
\hline
\multirow{2}{*}{API Endpoint}         & \multicolumn{1}{l|}{\multirow{2}{*}{Load Testing Scenario}} & \multicolumn{6}{c|}{Deployment Strategies Cost (in USD).}                                                                         \\ \cline{3-8} 
                                      & \multicolumn{1}{l|}{}                                       & \multicolumn{2}{c|}{GKE-50} & \multicolumn{2}{c|}{GKE-80} & \multicolumn{1}{c|}{OW}     & \multicolumn{1}{l|}{GCR} \\ \hline
\multirow{3}{*}{Sensors-Get}          & Linear                                                      & \multicolumn{2}{c|}{0.1054} & \multicolumn{2}{c|}{0.1053} & \multicolumn{1}{c|}{0.5195} & \textbf{0.0542}          \\ \cline{2-8} 
                                      & Random                                                      & \multicolumn{2}{c|}{0.1043} & \multicolumn{2}{c|}{1.0433} & \multicolumn{1}{c|}{0.4752} & \textbf{0.0544}          \\ \cline{2-8} 
                                      & Spike                                                       & \multicolumn{2}{c|}{0.4151} & \multicolumn{2}{c|}{0.4153} & \multicolumn{1}{c|}{1.2230} & \textbf{0.0586}          \\ \hline
\multirow{3}{*}{HTTP-Gateway}         & Linear                                                      & \multicolumn{2}{c|}{0.1225} & \multicolumn{2}{c|}{0.1276} & \multicolumn{1}{c|}{0.2592} & \textbf{0.0710}          \\ \cline{2-8} 
                                      & Random                                                      & \multicolumn{2}{c|}{0.1217} & \multicolumn{2}{c|}{0.1278} & \multicolumn{1}{c|}{0.4752} & \textbf{0.0630}          \\ \cline{2-8} 
                                      & Spike                                                       & \multicolumn{2}{c|}{0.4760} & \multicolumn{2}{c|}{0.4742} & \multicolumn{1}{c|}{1.0206} & \textbf{0.0539}          \\ \hline
\multirow{3}{*}{Users-Get}            & Linear                                                      & \multicolumn{2}{c|}{0.1103} & \multicolumn{2}{c|}{0.1113} & \multicolumn{1}{c|}{0.5401} & \textbf{0.0493}          \\ \cline{2-8} 
                                      & Random                                                      & \multicolumn{2}{c|}{0.1098} & \multicolumn{2}{c|}{0.1101} & \multicolumn{1}{c|}{0.3221} & \textbf{0.0478}          \\ \cline{2-8} 
                                      & Spike                                                       & \multicolumn{2}{c|}{0.4425} & \multicolumn{2}{c|}{0.4368} & \multicolumn{1}{c|}{1.3610} & \textbf{0.0475}          \\ \hline
\multirow{3}{*}{Devices-Add}          & Linear                                                      & \multicolumn{2}{c|}{0.1210} & \multicolumn{2}{c|}{0.1291} & \multicolumn{1}{c|}{0.4734} & \textbf{0.0617}          \\ \cline{2-8} 
                                      & Random                                                      & \multicolumn{2}{c|}{0.1178} & \multicolumn{2}{c|}{0.1295} & \multicolumn{1}{c|}{0.5020} & \textbf{0.0569}          \\ \cline{2-8} 
                                      & Spike                                                       & \multicolumn{2}{c|}{0.4506} & \multicolumn{2}{c|}{0.4630} & \multicolumn{1}{c|}{1.2797} & \textbf{0.0586}          \\ \hline
\multirow{3}{*}{Devices-Get}          & Linear                                                      & \multicolumn{2}{c|}{0.1081} & \multicolumn{2}{c|}{0.1087} & \multicolumn{1}{c|}{0.3677} & \textbf{0.0517}          \\ \cline{2-8} 
                                      & Random                                                      & \multicolumn{2}{c|}{0.1073} & \multicolumn{2}{c|}{0.1079} & \multicolumn{1}{c|}{0.5151} & \textbf{0.0521}          \\ \cline{2-8} 
                                      & Spike                                                       & \multicolumn{2}{c|}{0.4496} & \multicolumn{2}{c|}{0.4346} & \multicolumn{1}{c|}{1.3464} & \textbf{0.0586}          \\ \hline
\multirow{3}{*}{Consumer-Consume-Get} & Linear                                                      & \multicolumn{2}{c|}{0.1157} & \multicolumn{2}{c|}{0.1162} & \multicolumn{1}{c|}{0.2660} & \textbf{0.0636}          \\ \cline{2-8} 
                                      & Random                                                      & \multicolumn{2}{c|}{0.1135} & \multicolumn{2}{c|}{0.1141} & \multicolumn{1}{c|}{0.2636} & \textbf{0.0624}          \\ \cline{2-8} 
                                      & Spike                                                       & \multicolumn{2}{c|}{0.4432} & \multicolumn{2}{c|}{0.4439} & \multicolumn{1}{c|}{1.0527} & \textbf{0.0657}          \\ \hline
\end{tabular}
\end{adjustbox}

\caption{Cost in USD cents per 1000 requests for the different API endpoints across the different deployment strategies and load testing scenarios. The highlighted values represent the minimum costs for a particular workload.}
\label{tab:costcomparison}
\shrinkspace
\end{table}

\subsection{Comparing Costs}
\label{sec:costcomp}
Table~\ref{tab:costcomparison} shows the comparison between cost per 1000 requests for the different deployment strategies across the different workloads. For estimating the costs for GKE and OW, we use the computational model used by Google based on the VM type, the amount of persistent storage, the experiment duration, and the cluster management fee~\cite{gkepricing}. On the other hand for GCR, we use the number of invocations, the allocated memory, and the execution duration for estimating costs~\cite{gcrpricing}. Across all endpoints, we observe that GCR leads to the lowest cost per 1000 requests across all load testing scenarios.  This can be attributed to two reasons. First, GCR serves a high amount of requests per second, while at the same time it lacks expensive fixed costs. Second, the reservation-based fees of the GKE and OW setups are charged independently of the workload that actually occurs. For instance, for the \texttt{Users-Get} endpoint with the linear workload, the GKE-50 cluster would need to serve 2.24x number of requests (\~405,744) with the same execution duration to match the costs of GCR. 
However, invocation costs with GCR will not be cheaper if the request response times become significantly higher than the GKE and OW deployments. For these scenarios, the fixed cost structure benefits the other two deployment strategies.


\subsection{Lessons Learned}
From our migration and performance analysis, we highlight three lessons learned. First, FaaS and serverless CaaS technologies suffer from significantly high initial response times on a burst of function invocation requests due to function \textit{cold starts} as shown in Figure~\ref{fig:perfresults}. Although, mitigating cold starts is an active area of research in serverless computing~\cite{coldstarts, tppfaas}, with solutions such as pre-warming function instances before incoming requests on commercial cloud providers~\cite{awscoldstart}, it can lead to a significant number of application SLO violations if not accounted for by the developers. Second, the microservices based deployment strategies outperform the serverless deployment strategies for simple API endpoints responsible for fetching the required data from the database. For instance, for the \texttt{Users-Get} endpoint GKE-50 was $1.1$x faster than GCR and served $2682$ more requests. To this end, for better performance, the developers should consider using a microservices-based architecture for similar API calls that are invoked frequently and have a static response size. Third, the process of migrating a microservices-based application is mostly \textit{ad-hoc}, time consuming, and costly. As a result, developers should consider the percentage of existing code that can be resused as an important criterion before initiating the migration process (\S\ref{sec:methodology}). Moreover, language of the application and the choice of the serverless platform is also important. For instance, while GCR supports functions in any programming language, Google Cloud Functions (GCF) supports limited function runtimes with no support for languages such as \texttt{C++}.


\section{Related Work}
\label{sec:relatedwork}
Due to the rising popularity of serverless computing and its various advantages, there are increasing debates about the architecture design of modern cloud native applications when it comes to choosing microservices, serverless, or a hybrid approach. Towards this, some prior work~\cite{jindal2020, fan2020, jin2021} has focused on migrating and comparing the performance of microservices and serverless deployment strategies. Fan et al.~\cite{fan2020} migrate a simple employee time sheet management application onto AWS Lambda and compare its performance with Amazons Elastic Container Service. Jindal et al.~\cite{jindal2020} build on this and present comparison results for a simple cinema application across GKE, OW, and Google Cloud functions. In contrast to our work, the applications chosen for migration are relatively simple with only one \textit{off-the-shelf} software component. Jin et al.~\cite{jin2021} migrate four stateful microservices applications onto OW. The authors highlight lessons learnt from the migration and describe methods to minimize code changes while maintaining a similar performance. However, in contrast to our work they only present preliminary performance results. Moreover, none of the previous works present a cost comparison between the different deployment strategies.

\section{Conclusion \& Future Work} \label{sec:conclusion}
\label{sec:conclusion}

In this paper, we migrated a complex IoT platform application based on the microservices architecture onto OW and GCR. We comprehensively evaluated the performance of the application using different deployment strategies, i.e., GKE, OW, and GCR across different load testing workloads. From our experiments, we observed that GKE performed best for most scenarios followed by GCR and OW. However, using GCR led to least costs across all scenarios. In the future, we plan to migrate a suite of different complex microservices applications and provide detailed guidelines and performance measurements that will enable architects to make reasoned decisions about the architecture to use for their applications.

\section{Acknowledgement}
The research leading to these results was funded by the Deutsche Forschungsgemeinschaft (DFG, German Research Foundation)-Proje\-ktnummer 146371743-TRR 89: Invasive Computing. Google Cloud credits in this work were provided by the \textit{Google Cloud Research Credits} program with the award number 64c92de5-fb62-4386-8c5b-ff3f480390bb.

\bibliographystyle{ACM-Reference-Format}
\bibliography{serverless}


\end{document}